\documentclass[lettersize,journal]{IEEEtran}
\usepackage{amsmath,amsfonts}
\usepackage{algorithmic}
\usepackage{algorithm}
\usepackage{array}
\usepackage[caption=false,font=normalsize,labelfont=sf,textfont=sf]{subfig}
\usepackage{textcomp}
\usepackage{stfloats}
\usepackage{url}
\usepackage{xcolor}
\usepackage{verbatim}
\usepackage{graphicx}
\usepackage{cite}
\usepackage{booktabs} 
\usepackage{subfig}
\usepackage{array}    
\usepackage{amsmath}  
\usepackage{multirow}  
\usepackage[normalem]{ulem}
\usepackage[inline]{enumitem}
\begin{document}
\title{Sensing in Low-altitude Wireless Networks:
Systems, Techniques, and Developments}

    \author{\IEEEauthorblockN{Zihao Tao, Yiming Zhao, Hongtao Zhao, Zijun Gong, Ying Cui}
        \thanks{
        The authors are with the Internet of Things Thrust, The Hong Kong University of Science and Technology (Guangzhou), China (Corresponding author: Ying Cui). 
	}
	}
\maketitle
\begin{abstract}
The highly dynamic and safety-critical characteristics of low-altitude airspace render sensing an indispensable component of low-altitude wireless networks (LAWN). 
Although sensing techniques have been extensively studied under diverse paradigms, a prominent mismatch persists between state-of-the-art sensing schemes and the practical sensing demands of LAWN.
To fill this research gap,
this article systematically reviews LAWN-oriented sensing from the dimensions of system framework, core technologies, and research trends.
Specifically, we first analyze the sensing system framework, covering concepts, services and tasks, nodes and targets, and scenarios for LAWN sensing.
Next, we conduct a comparative analysis of existing sensing techniques from the perspectives of propagation medium, cooperation, methodology, and modality, analyzing their advantages and limitations.
Then, we summarize promising future research directions for deployable LAWN sensing systems, covering non-cooperative and cooperative
sensing, model-driven and data-driven sensing, and model-and-data-driven multi-modal sensing. Finally, we present a case study of a model-and-data-driven multi-modal method for real-time aerial target sensing. Compared with existing surveys on LAWN or sensing, this article delivers a more comprehensive, targeted review exclusively centered on LAWN sensing. 

\end{abstract}
\section{Introduction}
Low-altitude wireless networks (LAWNs) are widely recognized as an emerging networking paradigm that extends terrestrial communication infrastructure into low-altitude airspace (below 3,000 meters above ground level), forming a three-dimensional (3D), service-aware network fabric \cite{Jin2025_TVT}. LAWN constructs a unified, seamless, and intelligent integrated ground–air architecture. Its core objective is to enable the digitalization and intelligent management of low-altitude airspace through the joint optimization of communication, sensing, computing, control, and agentic artificial intelligence.  

Sensing in LAWN aims to provide comprehensive situation awareness, including target entity awareness and environmental awareness. In the highly dynamic and safety-critical low-altitude airspace, where communication, computing, and control are tightly coupled with environmental conditions, sensing is far more than an auxiliary capability but a core component of LAWN \cite{Jiang2025_ComMag}, providing a critical foundation for LAWN operations. Sensing in LAWN involves fundamental sensing services, including target sensing and environmental sensing, which underpin upper-layer sensing-enabled applications such as airspace management, communication coverage optimization, network resource scheduling, etc.  These basic sensing services are sustained by sensing tasks such as detection, estimation, tracking, identification, and mapping \cite{Yuhong_2025_Chinacom}.

LAWN comprises three categories of nodes: ground nodes (e.g., base stations (BSs), radar stations, roadside units (RSUs), etc.), aerial nodes (e.g., unmanned aerial vehicles (UAVs), electric vertical take-off and landing (eVTOL) aircraft, airships, etc.), and computation, control, and reference (CCR) nodes (e.g., cloud/edge servers, flight control units (FCUs), real-time kinematic (RTK) stations, etc.). Sensing targets within LAWN fall into two groups: aerial targets (e.g., aircraft, flying animals, drifting objects, etc.) and ground targets (e.g., motor vehicles, pedestrians, cyclists, trees, etc.). 
Sensing scenarios in LAWN are generally divided into three typical types, namely air-to-ground (A2G), air-to-air (A2A), and ground-to-air (G2A). These scenarios confront a set of common technical challenges: rapid three-dimensional mobility of aerial nodes and targets, including high dynamic 3D mobility of aerial nodes and targets, insufficient illumination and adverse weather conditions, dense distribution of targets, weak echoes from small radar cross section (RCS) targets, and intensive near-ground clutter and multipath effects.

Existing sensing techniques can be classified by four criteria (propagation medium, cooperation mode, modeling methodology, and sensing modality), with each criterion corresponding to two distinct subcategories: RF/optical-based, non-cooperative/cooperative, model-driven/data-driven, and single/multi-modal sensing. 
 Specifically, RF-based sensing adapts to varying lighting and weather conditions, long sensing range, and high-speed targets but suffers from clutter and multipath effects. In contrast, optical-based sensing provides precise spatial perception and resistance to clutter and multipath effects but is susceptible to insufficient lighting and adverse weather conditions, long sensing ranges, and high-speed targets. Non-cooperative sensing enables lightweight independent sensing but suffers from occlusion, target glint ambiguity, and weak echoes from small-RCS targets. In contrast, cooperative sensing enables multi-node complementary sensing but suffers from rigid scheduling, time-frequency misalignment, high signaling and computation overhead, and scalability issues. Model-driven sensing possesses clear physical interpretability but suffers from inaccurate RF/optical modeling and a lack of 3D modeling for aerial sensing. In contrast, data-driven sensing possesses strong environmental adaptability but suffers from low physical interpretability and insufficient generalization capability. Single-modal sensing enables lightweight deployment but suffers from inherent modal limitations. In contrast, multi-modal sensing enables cross-modal complementary fusion but suffers from spatiotemporal misalignment, cross-modal feature imbalance, and excessive computational overhead.

These sensing techniques may exhibit degraded performance when deployed in LAWN because of the aforementioned limitations.
Therefore, in-depth research tailored to LAWN sensing is urgently required. 

Sensing-related content has been briefly covered in existing surveys concerning ISAC \cite{ISACSurvey_Wu_ComMag_2024,UAVSurvey_Zhang_ComMag_2024,Liang_CIOTSC_2025} and LAWN \cite{LAWNSurvey_Jun_Chinacom2026,LAWNSurvey_Jia_ComMag_2026}, yet prominent research gaps targeting LAWN-native sensing remain largely unaddressed.  
First, prior reviews illustrate separate system architectures for generic ISAC and LAWN, while none establish a dedicated sensing framework customized for LAWN that systematically elaborates sensing concepts, supporting services, core tasks, network nodes, ground and aerial targets, and typical operation scenarios.   
Second, some surveys briefly discuss RF/optical-based, cooperative/non-cooperative, conventional/AI-enhanced, and single/multi-modal sensing techniques, but none comprehensively analyze their features, advantages, and limitations. Finally, several surveys outline prospective research trends for general ISAC or LAWN networks, yet few concentrate on specialized technical solutions oriented toward LAWN sensing.

To address these issues, this article systematically reviews LAWN-oriented sensing from the dimensions of system framework, core technologies, and research trends. Compared with prior surveys on generic LAWN or universal sensing, this article focuses exclusively on LAWN-specific sensing. The contributions are summarized as follows.
\begin{enumerate*}
\item  We first analyze the sensing system framework,  covering fundamental concepts, sensing services and tasks, network nodes and perception targets, and typical scenarios for LAWN sensing.
\item We conduct a comparative analysis of existing sensing techniques from the perspectives of propagation medium, cooperation, methodology, and modality and elaborate on their inherent merits and technical bottlenecks.
\item We summarize promising future research directions for deployable LAWN sensing systems,  covering non-cooperative/cooperative sensing, model-driven/data-driven sensing, and model-and-data-driven multi-modal sensing.
\item We present a case study of a model-and-data-driven multi-modal method for real-time aerial target sensing. 
\end{enumerate*}  
Therefore, compared with prior surveys, this article delivers a more comprehensive, targeted review exclusively centered on LAWN sensing.

\section{Sensing System Framework in LAWN}
\label{section:service}
In this section, we introduce the sensing system framework in LAWN, including sensing concepts, services,  tasks, nodes, targets, and scenarios.

\subsection{Sensing Concepts, Services, and Tasks}

Sensing in LAWN aims to provide comprehensive situation awareness, including target entity awareness and environmental awareness, to support LAWN operations. It involves basic sensing services such as target sensing and environmental sensing, which form the basis for other sensing-related services, such as airspace management (e.g., collision avoidance and dynamic trajectory planning and control), communication coverage optimization, network resource scheduling, abnormal target early warning, flight safety assessment, multi-node cooperative maneuvering, etc. These basic sensing services are supported by basic sensing tasks, including detection, estimation,  tracking, prediction, identification, and mapping \cite{Yuhong_2025_Chinacom}.

Target sensing focuses on maintaining continuous awareness of aerial or ground targets within designated low-altitude regions, supporting other LAWN services such as airspace management, abnormal target early warning, multi-node cooperative maneuvering, etc. This service relies on target detection (over 95\% detection accuracy with a lower than 0.5\% false alarm rate), target localization (decimeter-level accuracy), target tracking (below 1\% tracking loss rate), target trajectory prediction (below 10 ms), and target identification (over 90\% multi-class recognition accuracy).

Environmental sensing focuses on characterizing surrounding conditions that influence low-altitude operations, such as terrain structures, ground building layouts, static and dynamic obstacles, meteorological changes, and wind field disturbances, supporting other LAWN services such as  communication coverage optimization, network resource scheduling, flight safety assessment, etc. This service relies on spectrum estimation (100 kHz-level resolution), wind velocity estimation (estimation error below 0.5 m/s), spatial mapping of terrain and obstacles (meter-level resolution), etc.

\subsection{Nodes and Targets}

\begin{figure}[t] 
  \centering
  \small
\includegraphics[width=0.48\textwidth]{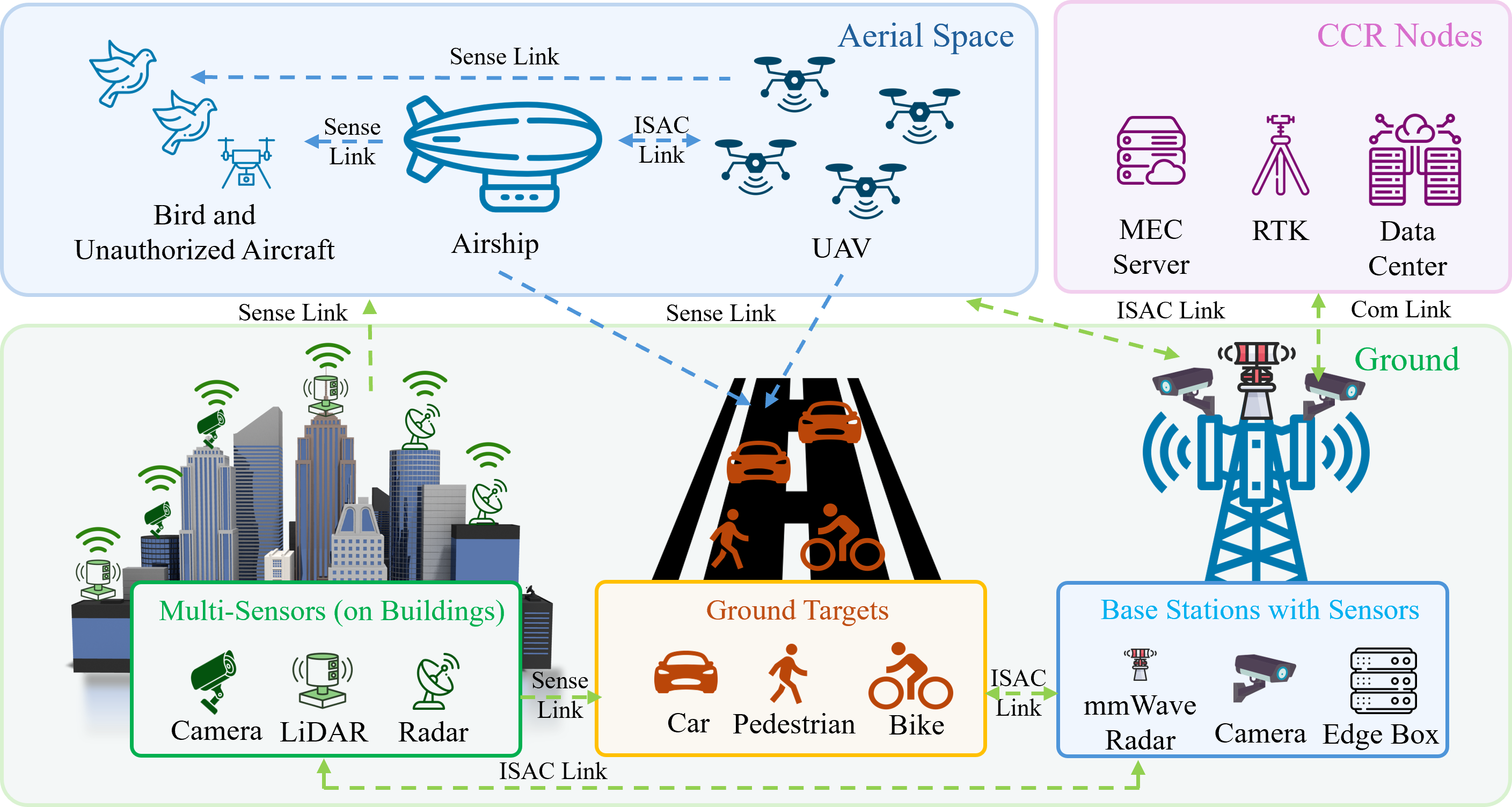}
  \vspace{-0.1in}
\caption{Overview of LAWN system.
}
\label{fig:system}
\vspace{-0.2in}
\end{figure}

As illustrated in Fig.~\ref{fig:system}, LAWN is composed of aerial nodes, ground nodes, and CCR nodes. A summary of these nodes is shown in Table~\ref{tab:lawn}. 
Aerial nodes are aircraft with multi-modal sensors, including the RGB/IR camera, mmWave radar, LiDAR, ultrasonic ranging sensor, inertial measurement unit (IMU), and altimeter. They possess 3D mobility and adjustable observation perspectives.
Ground nodes are ground infrastructure such as RGB/IR camera stations, microwave/mmWave radar stations, LiDAR stations, BSs, RSU stations, meteorological sensor stations, thermal imaging stations, and compact monitoring stations on the ground, rooftops, or towers. In contrast, they have fixed positions, inevitably constraining their observation perspectives.

Both aerial and ground nodes can be divided into two types: resource-constrained nodes and resource-sufficient nodes.
Resource-constrained aerial nodes, such as UAVs and eVTOLs, and resource-constrained ground nodes, such as compact monitoring stations, are equipped with low-power compact sensor modules and system-on-chips (SoCs), realizing short-range, low-complexity standalone sensing. Complex standalone sensing and cooperative sensing have to be offloaded to edge or cloud servers.
Resource-sufficient aerial nodes, such as airships, tethered balloons, and helicopters, and ground nodes, such as BSs, radar stations, and RSUs, are equipped with high-performance sensor modules and FPGAs/CPUs, supporting 
long-range, low/medium/high-complexity standalone sensing.
Cooperative sensing has to be offloaded to edge or cloud servers.

CCR nodes refer to computation nodes, control nodes, and reference nodes, responsible for core backend scheduling and service support. Computation nodes include cloud and edge servers, undertaking information processing, situation analysis, and route planning. Control nodes include cloud and edge servers and FCUs, realizing hierarchical management from global scheduling to onboard execution. Reference nodes include global navigation satellite system (GNSS), RTK, time, and automatic dependent surveillance-broadcast (ADS-B) reference stations, providing unified positioning, time, and air situation benchmarks for the whole system.

\newsavebox{\uppertablebox}
\newlength{\uppertablewidth}

\begin{table*}[t]
\centering
\caption{Nodes and Targets in LAWN}
\vspace{-0.1in}
\label{tab:lawn}
\renewcommand{\arraystretch}{1.06}
\setlength{\tabcolsep}{4.8pt}
\savebox{\uppertablebox}{%
  \begin{tabular}{p{1.6cm} p{6.7cm} p{5cm} p{4cm}}
  \textbf{Node Class} & \textbf{Node type \& placement} & \textbf{Sensors} & \textbf{{Resources}} \\
  \hline
  Aerial nodes &
   Aircraft with multi-modal sensors; onboard such as UAVs, eVTOLs, helicopters, lighter‑than‑air platforms (e.g., airships and balloons) &
  RGB/IR camera, mmWave radar, LiDAR, IMU, {ultrasonic ranging sensors, 5G cellular module, altimeter} &
  \multirow{2}{=}{Resource-constrained: low-power compact sensor modules and SoCs.\\ Resource-sufficient: high-performance sensor modules and FPGAs/CPUs} \\
  \cline{1-3}
  Ground nodes &
  BSs, microwave radar/mmWave radar/RSUs/RGB camera/IR camera stations, compact monitoring stations; ground/rooftops/towers &
   BSs, microwave/mmWave radars, RSUs, RGB/IR cameras, meteorological sensors, thermal imagers, LiDARs & \\
  \hline
  CCR nodes &
  Cloud/edge servers, FCUs, GNSS/RTK/time/ADS-B reference stations; data center &
  GNSS/RTK &
  High computation performance for global fusion and calibration \\
  \hline
  \end{tabular}%
}
\settowidth{\uppertablewidth}{\usebox{\uppertablebox}}

\scalebox{0.85}{
  \begin{tabular}{@{}c@{}}   
  \toprule
  \usebox{\uppertablebox}     
  \\[-2pt]
  \midrule                   
\resizebox{450pt}{!}{%
    \setlength{\tabcolsep}{3pt}
    \renewcommand{\arraystretch}{0.8}
    \scriptsize
    \begin{tabular}{ccccccccc}
    \textbf{Target Attribute} & \textbf{Pedestrians} & \textbf{Cyclists} & \textbf{Cars} & \textbf{Trucks} & \textbf{UAVs} & \textbf{eVTOL} & \textbf{Airships} & \textbf{Birds} \\
    \midrule
    Material & Biological & Biological/Metal & Metal & Metal & Plastic & Metal/Composite & Synthetic/Plastic & Biological \\
    RCS (m$^2$) & 0.1--0.5 & 0.1--0.6 & 0.1--1 & 0.5--2 & 0.001--0.01 & 0.1--1 & 0.1--1 & 0.001--0.01 \\
    Vertical Velocity (m/s) & 0 & 0 & 0 & 0 & 5--10 & 0--8 & 0--5 & 5--15 \\
    Horizontal Velocity (m/s) & 1.5 & 4--8 & 10--20 & 8--15 & 10--30 & 40--70 & 0--5 & 10--20 
    \end{tabular}%
  }
  \\[-2pt]
  \bottomrule 
  \end{tabular}%
}
\vspace{-0.2in}
\end{table*}

 Targets in LAWN include aerial and ground targets. A summary of representative targets is shown in Table~\ref{tab:lawn}.
 Aerial targets mainly include authorized and unauthorized aircraft, flying animals, and drifting objects, possessing flexible 3D movement unaffected by terrain.
Ground targets mainly include motor vehicles (e.g., cars and trucks), pedestrians, cyclists, buildings, and trees, as well as other static and dynamic ground entities, possessing limited 2D movement constrained by terrain. Moreover, aerial and ground targets may have
different sizes and materials and appear in groups of distinct densities.

\subsection{Sensing Scenarios}
Sensing scenarios in LAWN are broadly categorized into three types, i.e., aerial nodes sensing ground targets, ground nodes sensing aerial targets, and aerial nodes sensing other aerial targets, referred to as A2G, G2A, and A2A, respectively, as shown in Fig.~\ref{fig:system}.
These three scenarios share a range of common challenges.
First, since both aerial nodes and targets possess flexible 3D mobility, sensing in all three scenarios needs to handle their dynamic movements.
Second, since RGB/IR cameras and LiDAR modalities suffer from poor robustness against insufficient lighting and harsh weather conditions, the resulting sensing performance remains unstable across the three scenarios.
Third, the dense distribution of aerial and ground targets poses great challenges to multi-target detection across the three scenarios.
Fourth, since most aerial targets, such as UAVs, eVTOLs, flying creatures, and aerial debris, have small RCS and produce weak RF signal echoes, A2A and G2A sensing face notable difficulties.
Finally, near-ground environments feature severe ground clutter, prominent multipath effects, and diverse terrain obstacles, which significantly degrade the RF‑based sensing performance in A2G and G2A scenarios.

\begin{table*}[t]
\centering
\caption{Comparison Between Model- and Data-driven Sensing}
\vspace{-0.1in}
\label{tab:ai_vs_conventional}
\renewcommand{\arraystretch}{1.15}
\scalebox{0.85}{
\begin{tabular}{lcc}
\hline
\textbf{Aspect} & \textbf{Model-driven Sensing} & \textbf{Data-driven Sensing} \\
\hline
Signal Model Requirement 
& Requires explicit analytical models 
& Can operate with implicit or unknown models \\

Robustness to Model Mismatch 
& Sensitive to mismatch 
& More adaptive to environment variations \\

Data Requirement 
& Limited training data required 
& Requires large datasets \\

Computational Complexity 
& Moderate and predictable 
& Potentially high (training phase) \\

Interpretability 
& High (theory-grounded) 
& Often limited (black-box behavior) \\

Generalization Capability 
& Scenario-specific tuning required 
& Better cross-scenario adaptability \\

Typical Methods 
& CFAR, MUSIC, MLE, Kalman filtering 
& RNN, CNN, Transformer, GNN \\

Suitable LAWN Scenarios 
& Structured and well-modeled environments 
& Complex, dynamic, heterogeneous environments \\
\hline
\end{tabular}}
\vspace{-0.15in}
\end{table*}

\begin{table*}[t]
\centering
\caption{Sensing Modalities in LAWN}
\vspace{-0.1in}
\label{tab:lawn_modalities_pros_cons_clean}
\renewcommand{\arraystretch}{1.06}
\setlength{\tabcolsep}{4.2pt}
\scalebox{0.85}{
\begin{tabular}{p{1.3cm} p{2.8cm} p{1.5cm} p{5.2cm} p{5.0cm}}
\hline
\textbf{Modality} & \textbf{Outputs} & \textbf{Max range} & \textbf{Strengths} & \textbf{Limitations} \\
\hline
ISAC/ mmWave Radar &
Range, angle, velocity, micro-Doppler &
0.2 - 2 km/ 50 - 300 m &
Low-cost deployment, reliable tracking of high-speed targets, robust to weather and illumination, medium to long-range sensing &
Sensitive to clutter and multipath effects, insensitive to weak echoes of small-RCS targets, degraded by occlusion \\
\hline
Camera &
Texture, shape, category, depth via inference, velocity via tracking &
20 - 100 m &
Low-cost deployment, precise spatial perception, fine-grained detail recognition, 
appearance-rich cues &
Sensitive to illumination and weather, depth and velocity indirect, degraded by motion blur and occlusion \\
\hline
LiDAR &
3D points, depth, intensity, size and shape cues &
50 - 300 m &
Dense 3D structure, high-precision localization, resistance to ground clutter and multipath effects &
Power and compute intensive, sparse at long range, degraded in fog, rain, and snow \\
\hline
\end{tabular}}
\vspace{-0.2in}
\end{table*}

\section{Existing Sensing Techniques}
\label{section:techniques}
In this section, we review existing sensing techniques in terms of propagation medium, cooperation, methodology, and modality and identify their advantages and limitations.

\subsection{{RF-based versus Optical-based Sensing}}
\label{section_medium}

Existing sensing techniques are categorized into RF-based and optical-based sensing based on the propagation medium. Both have been studied for all three scenarios in LAWN. 
\subsubsection{RF-based Sensing} RF-based sensing extracts perception information by analyzing the amplitude, phase, and Doppler shift of received RF signals.
It includes microwave/mmWave radar-based sensing \cite{Li_colocatedMIMOradar2007} and communication-assisted sensing (or ISAC \cite{Barneto_FDISAC_TMTT2019}). Its major advantages lie in strong penetration capability, long sensing range, reliable tracking of high-speed targets, and robustness against varying lighting conditions. However, it suffers from small RCS, ground clutter, and multipath effects that severely distort echo signals, which degrade detection and estimation accuracy in complex scenarios.
\subsubsection{Optical-based Sensing}
Optical-based sensing extracts perception information by analyzing the intensity, texture, and spatial features of received optical signals. It contains 
RGB/IR camera-based and LiDAR-based sensing. Its major advantages lie in precise spatial perception, fine-grained detail recognition, and strong resistance to ground clutter and multipath effects.
However, it faces inherent drawbacks including limited detection range, unstable tracking of high-speed targets, insufficient lighting conditions, and harsh weather conditions (such as dense fog and heavy snow) that drastically attenuate optical signals, resulting in degraded sensing continuity and poor environmental adaptability in complex outdoor scenarios.

\subsection{Non-cooperative versus Cooperative Sensing}

Existing sensing techniques can be classified into non-cooperative and cooperative sensing from the perspective of node interaction modes. Non-cooperative sensing has been studied for all three  LAWN sensing scenarios, while cooperative sensing is mainly explored in A2G and A2A scenarios.

\subsubsection{Non-cooperative Sensing}
\label{section_noncooperative}

Non-cooperative sensing extracts perception information by relying solely on individual nodes to perform independent perception using only local observation data\cite{Hochberg2026_TRS,He_FDISAC_JSAC2023}. Its major advantages lie in low system complexity, flexible and lightweight deployment, and minimal signaling overhead without inter-node interaction. 
However, multiple inherent limitations remain. First, non-cooperative sensing is susceptible to spatial occlusion, yielding discontinuous sensing coverage. Second, non-cooperative sensing induces severe target ambiguity originating from the glint effect. Third, non-cooperative sensing lacks multi-station joint time-frequency integration gains, thus degrading the detection performance of small-RCS targets.

\subsubsection{Cooperative Sensing}

Cooperative sensing realizes comprehensive joint perception in centralized or decentralized architectures by exploiting multiple distributed collaborative nodes, local observations, and intermediate sensing results \cite{Romero2013_TAES}. Its prominent strengths include enhanced observation diversity, strong robustness against spatial occlusion, and substantially improved sensing performance via multi-node complementary perception that compensates for single-view observation limitations. Nevertheless, it still suffers from critical inherent drawbacks. First, cooperative sensing relies predominantly on ad-hoc or fixed collaboration mechanisms, lacking effective scheduling and adaptability. Second, insufficient inter-node time-frequency synchronization in cooperative sensing induces spatiotemporal misalignment across distributed nodes and degrades overall sensing performance. Third, frequent information exchange among distributed nodes or between them and a central node generates excessive signaling overhead and intensifies contention over limited communication resources,  introducing extra transmission latency. 
Fourth, intensive computing workloads at distributed or central nodes consume massive computational resources and induce excessive computation latency.   
Fifth, the joint constraints of scarce communication, computing, and energy resources severely restrict the system scalability of cooperative sensing. 

\subsection{Model-driven versus Data-driven Sensing}
Existing sensing techniques can be classified into model-driven sensing and data-driven sensing based on processing methodology. Both have been investigated for all three LAWN scenarios. In the sequel, we exclusively discuss these two within the scope of single-modal sensing for ease of illustration.

\subsubsection{Model-driven Sensing}
Model-driven sensing extracts perception information using analytical physical signal models \cite{Hochberg2026_TRS,He_FDISAC_JSAC2023}, covering both RF-based and optical-based sensing paradigms. Specifically, representative RF-based sensing methods include matched filtering and CFAR-based detection, maximum-likelihood and subspace-based estimation, micro-Doppler-based identification, and Kalman and particle filtering-based tracking, while typical optical-based sensing methods consist of image threshold segmentation, edge detection, optical flow tracking, LiDAR point cloud geometric fitting, and clustering filtering algorithms. The major advantages of model-driven sensing lie in explicit physical interpretability, yielding stable sensing performance under ideal model-matched propagation conditions, and low computational overhead.  
However, model-driven sensing has several limitations. First, RF-based sensing usually ignores multipath effects, self-interference, and clutter, mismatching real electromagnetic propagation characteristics. Second, RF-based sensing is often built upon 2D planar assumptions tailored for large-RCS ground targets, lacking effective modeling support for small-RCS aerial targets in 3D airspace. Third, optical-based sensing adopts oversimplified physical hypotheses for complex atmospheric and illumination disturbances, failing to reconstruct real-world optical transmission rules precisely.

\subsubsection{Data-driven Sensing}

Data-driven sensing extracts perception information from massive raw sensing data without relying on explicit physical signal propagation models \cite{Luong2026_CommunSurvTutor}, covering both RF-based and optical-based data-driven sensing paradigms. Specifically, representative RF-based methods include CNN-based target detection, Transformer-based parameter estimation, CNN-based target recognition, and RNN-based target tracking. In contrast, typical optical-based methods consist of CNN-based semantic segmentation, Transformer-based target detection, FlowNet-based motion tracking, and PointCNN-based target clustering.  
Its major advantages lie in strong adaptability to practical, complex environments. However, data-driven sensing exhibits poor physical interpretability as a black-box learning paradigm, lacking explicit physical and mathematical derivation support for sensing results. In addition, it suffers from limited generalization capability due to data scarcity, label noise, and domain distribution drift, making it difficult to adapt to various sensing scenarios.

\subsection{Single-modal versus Multi-modal Sensing}
Existing sensing techniques can be classified into single-modal and multi-modal sensing in terms of sensing modality. Single-modal sensing has been extensively investigated across all three LAWN sensing scenarios compared with multi-modal sensing.

\subsubsection{Single-modal Sensing}
Single-modal sensing extracts perception information relying on a single independent sensing modality. It includes microwave/mmWave radar-based sensing, ISAC-based sensing, RGB/IR camera-based sensing, and LiDAR-based sensing. Its major advantages lie in simple and low-cost system deployment and low computational overhead, requiring minimal hardware configuration and data processing capability to support lightweight basic sensing. However, RF-based modalities suffer from severe echo distortion caused by clutter and multipath effects, and optical modalities are vulnerable to signal attenuation under adverse weather and lighting conditions. Thus, single-modal sensing is constrained by the inherent defects of individual sensing modalities, lacking complementary cross-modal perception gains.

\subsubsection{Multi-modal Sensing}
Multi-modal sensing extracts comprehensive perceptual information by fusing multiple complementary sensing modalities \cite{Wei_IV2025}. Its prominent strengths lie in superior environmental robustness and holistic perception capability, which resist echo distortion, signal attenuation, and diverse interferences under complex field scenarios. Nevertheless, its fusion strategies have critical limitations. 
First, mainstream multi-modal fusion pipelines adopt naive direct data concatenation and neglect node heterogeneity and time-varying network dynamics, thereby triggering severe spatiotemporal misalignment among asynchronous multi-modal data streams. Second, existing fusion architectures overlook inherent inter-modal feature discrepancies and imbalanced information density, inducing feature suppression and skewed modal weight contributions during joint inference. Third, most fusion methods prioritize sensing accuracy at the cost of massive computational overhead, which exceeds the onboard computing and power budgets of some aerial platforms, incurring excessive inference latency and shortening sustainable flight endurance.

\section{{Future Research Directions}}
\label{section:challenge}

In this section, we outline future research directions for LAWN sensing, covering non-cooperative and cooperative sensing, model-driven and data-driven sensing, and model-and-data-driven multi-modal sensing.

\subsection{Non-cooperative Sensing}
To address the above-mentioned drawbacks of non-cooperative sensing, including severe occlusion vulnerability, target measurement ambiguity caused by target glint, and poor sensing performance against small-RCS targets, several promising future research directions are proposed. First, future work may investigate occlusion-robust temporal accumulation sensing schemes, leveraging continuous sequential echoes to recover incomplete target scattering signatures and improve anti-occlusion performance. 
Second, future research can explore multi-frame feature decoupling and glint suppression algorithms to eliminate measurement bias and stabilize sensing accuracy. Third, future studies may develop wideband high-gain beamforming and long-time coherent integration strategies to approach coherent accumulation gains and boost sensing performance for small-RCS targets.

\subsection{Cooperative Sensing}
To address the aforementioned bottlenecks of cooperative sensing, including rigid scheduling, time-frequency misalignment, excessive signaling overhead, intensive computing burdens, and limited system scalability, several promising research directions are put forward.  
First, future work may investigate adaptive collaboration and dynamic scheduling, enabling seamless adaptation to time-varying network topologies and heterogeneous task demands while fully tapping the spatial complementary gains of cooperative sensing. 
Second, future research can explore robust high-precision time-frequency synchronization and spatiotemporal alignment methods to mitigate clock drift and inter-node observation deviation, eliminating cross-node misalignment and further lifting overall cooperative sensing accuracy. 
Third, future studies may develop lightweight periodic interaction protocols coupled with spectrum-aware adaptive transmission policies to cut redundant inter-node signaling, ease wireless resource contention, and suppress excessive transmission latency during distributed sensing data exchange. Fourth, future work can design adaptive task partitioning and partial task offloading to balance computation and communication resource utilization among distributed and central nodes, reducing end-to-end sensing latency. 
Fifth, future research may construct hierarchical clustered cooperative architectures with intra-cluster local fusion to sustain system scalability under constrained communication, computing, and energy budgets.

\subsection{Model-driven Sensing}
To address the aforementioned limitations in model-driven sensing, such as severe RF model mismatch under complex propagation environments, insufficient 3D modeling and dedicated algorithm design for small-RCS aerial targets, and inaccurate optical modeling accounting for atmospheric and illumination disturbances, several promising research directions are put forward. First, future work may construct RF signal propagation models that fully characterize self-interference, time-varying multipath effects, and non-stationary ground clutter in complex near-ground scenarios and develop environment-adaptive sensing algorithms to realize authentic, reliable perception consistent with practical complex environmental features. Second, future research can develop targeted 3D spatial modeling and robust RF-based sensing algorithms integrated with long-time coherent integration customized for small-RCS aerial targets to achieve precise sensing within complex 3D airspace. Third, future studies may build high-fidelity optical physical models to accurately quantify illumination and atmospheric interference and embed self-adaptive parameter calibration modules to diminish modeling biases, sustaining reliable optical sensing performance in complex environments.

\subsection{Data-driven Sensing}
To address the aforementioned shortcomings in data-driven sensing, including low physical interpretability and insufficient generalization capability, several promising research directions are suggested. First, future work may investigate hybrid model-data-driven sensing paradigms embedded with physical governing equations and explicit physical-mathematical priors to mitigate physically uninterpretable black-box limitations and improve generalization against data scarcity and domain distribution drift. Second, future research can explore universal foundation signal sensing models with lightweight LoRA/adapter fine-tuning to extract shared multi-source signal features to adapt to diverse sensing scenarios under limited labels and enhance robustness against label noise and cross-sensing-domain discrepancies. Third, future studies may develop physics-informed meta-learning frameworks for sensing to extract invariant physical meta-knowledge across diverse measurement environments, realize fast zero/few-shot adaptation to unseen tasks, and align model outputs with interpretable physical mechanisms to jointly strengthen out-of-distribution generalization and physical interpretability.

\subsection{Model-and-Data-Driven Multi-modal Sensing}
To address the aforementioned challenges in multi-modal sensing, including multi-modal spatiotemporal misalignment, cross-modal feature imbalance, and excessive computational overhead, a unified model-and-data-driven multi-modal sensing framework is proposed, and several promising research directions are put forward. 
First, future work may design latency-aware fusion architectures and motion-compensated alignment algorithms to eliminate temporal and spatial mismatches of asynchronous multi-modal data streams in dynamic network environments. 
Second, future work may develop modality-aware balanced fusion mechanisms embedded with reliability evaluation modules to avoid feature suppression and fully exploit complementary sensing gains among heterogeneous modalities.
Third, future studies may explore lightweight compact fusion models and incremental data processing paradigms to reduce computational costs and satisfy real-time latency constraints on resource-constrained aerial sensing nodes.

\section{Case Study: Model-and-Data-Driven Multi-Modal Sensing}
\label{section:casestudy}

\begin{figure}[t]
\centering
\includegraphics[width=0.42\textwidth]{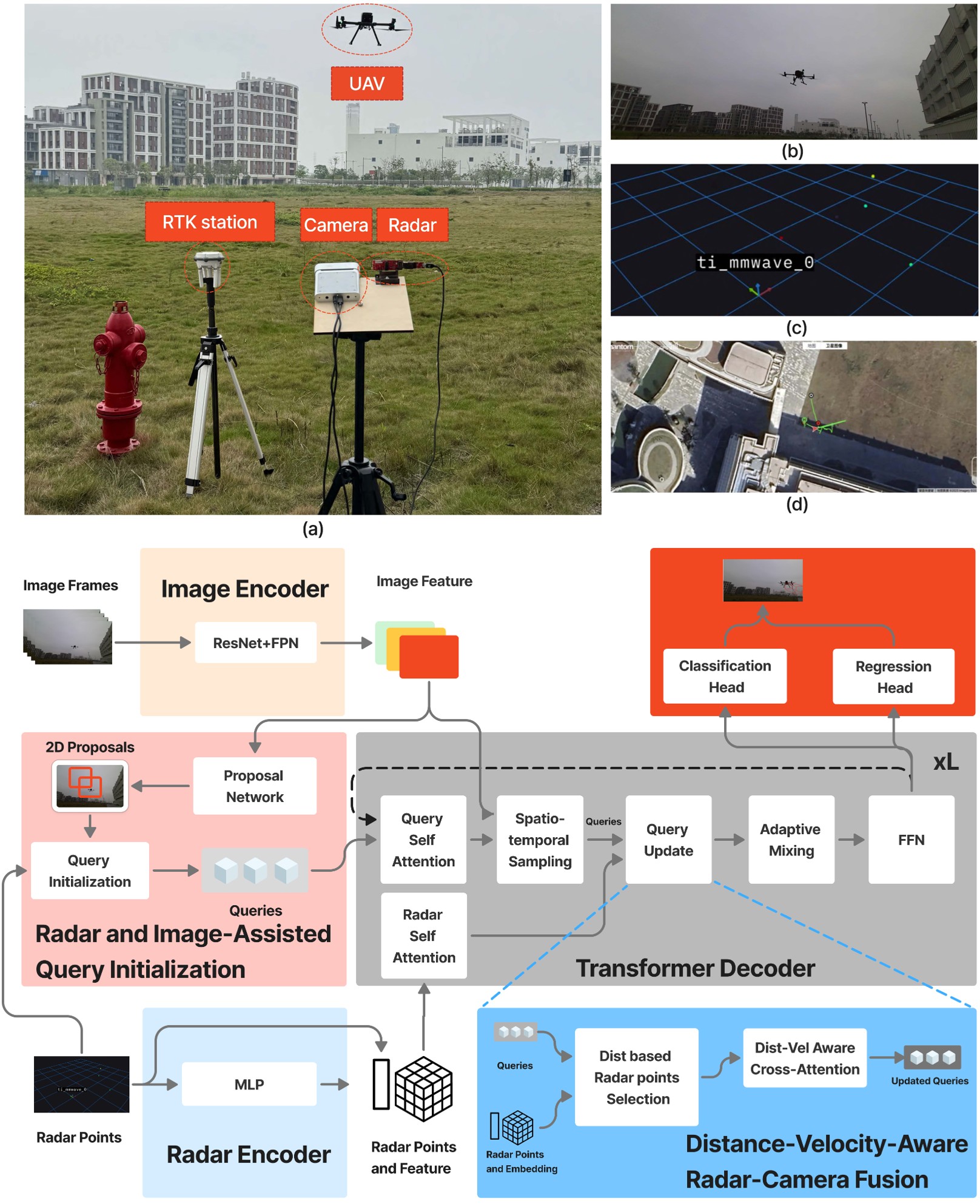}
\vspace{-0.1in}
\caption{Real-world radar-camera data collection platform and sparse radar-camera fusion method for UAV detection and estimation.}
\label{fig:casestudy_platform}
\vspace{-0.1in}
\end{figure}

To demonstrate the practical potential of model-and-data-driven multi-modal sensing in LAWN, we present a real-world case study on real-time 3D UAV detection and estimation using sparse radar-camera fusion. 
As shown in Fig.~\ref{fig:casestudy_platform}, a ground-deployed sensing platform is constructed to collect real-world multi-modal data and reference data by integrating a mmWave radar, an RGB camera, and an RTK-assisted UAV tracking system. Specifically, the radar provides 3D point clouds and velocity measurements, the camera provides appearance-rich visual observations, and the RTK system provides centimeter-level UAV position references for annotation and evaluation. The collected data covers different UAV types, outdoor environments, and lighting conditions, providing a practical basis for detecting UAV targets with small RCS and flexible 3D mobility in G2A sensing scenarios. 

Based on collected data, we develop a sparse radar-camera fusion method, as illustrated in Fig.~\ref{fig:casestudy_platform}. Its brief illustration is given below. Image features are extracted from RGB frames, while radar points are encoded by a lightweight radar encoder. Instead of constructing dense BEV representations, object queries are initialized using both image proposals and radar points. They are then refined through a distance-and-velocity-aware fusion mechanism, where radar Doppler and spatial-distance cues help associate sparse radar points with visual object queries. The proposed method represents a typical example of model-and-data-driven multi-modal sensing. We consider the following baselines. For image-only 3D detection, we employ well-known BEVFormer and SparseBEV as dense and sparse BEV detection baselines, respectively. We also compare our query coordinate initialization method with ray-centric query initialization. For radar-camera fusion detection, we select RaCFormer and RCM-Fusion covering different fusion architectures, and further compare our query update module with a range-adaptive radar aggregation strategy.

As shown in Table~\ref {tab:casestudy_results}, the sparse radar-camera fusion method achieves 83.20\% mean average precision (mAP) for detection and an absolute trajectory error (ATE) of 0.317 for trajectory estimation, with an inference latency of only 17 ms, outperforming image-only methods and existing radar-camera fusion baselines under the same tiny-model setting. This case study suggests that real-world complementary radar-camera observations and sparse fusion designs are essential for accurate, real-time UAV sensing in practical LAWN systems.
We also validate our design under different UAV types and light conditions. 
For drone type, the larger M350 RTK achieves 95.4\% mAP with lower localization and yaw errors than the smaller Mavic 3E, benefiting from a larger RCS. 
Under different lighting, all metrics degrade at night, with mAP dropping by over 20\% and ATE tripling, highlighting the difficulty of low-light UAV detection.

\begin{table}[t]
\centering
\caption{Case Study Results for Real-time Radar-camera UAV Sensing}
\vspace{-0.1in}
\label{tab:casestudy_results}
\renewcommand{\arraystretch}{1.05}
\setlength{\tabcolsep}{4pt}
\footnotesize
\scalebox{0.9}{
\begin{tabular}{lcccc}
\hline
\textbf{Method} & \textbf{Modality} & \textbf{mAP (\%)$\uparrow$} & \textbf{ATE$\downarrow$} & \textbf{Latency (ms)$\downarrow$} \\
\hline
BEVFormer-tiny & Camera & 70.10 & 0.468 & 21 \\
SparseBEV-tiny & Camera & 74.18 & 0.356 & 19 \\
RCM-tiny & Radar+Camera & 74.98 & 0.408 & 30 \\
RaCFormer-tiny & Radar+Camera & 76.32 & 0.333 & 32 \\
\textbf{SRCF-UAV-tiny} & Radar+Camera & \textbf{83.20} & \textbf{0.317} & \textbf{17} \\
\hline
\end{tabular}}
\vspace{-0.2in}
\end{table}

\section{Conclusion}
\label{section:conclusion}

This article provides a comprehensive overview of LAWN-specific sensing. We first dissect the core framework of LAWN sensing, including fundamental concepts, sensing services and tasks, network nodes, and perception targets, as well as typical sensing scenarios. Then, we conduct a comparative analysis of existing sensing techniques from the perspectives of propagation medium, cooperation mode, modeling methodology, and sensing modality and summarize their inherent merits and technical bottlenecks. Next, we outline promising future research directions for deployable LAWN sensing systems, covering non-cooperative/cooperative sensing, model-driven/data-driven sensing, and integrated model-and-data-driven multi-modal sensing. Finally, we present a case study of a model-and-data-driven multi-modal method for real-time aerial target perception.

\section{Acknowledgment}

This work was supported in part by the National Science and Technology Major Project of China on Mobile Information Networks under Grant 2024ZD1300400, the National Natural Science Foundation of China under Grants 62371412 and 62571467, the Guangdong Basic and Applied Basic Research Natural Science Funding Scheme under Grant 2024A1515011184, and the Guangzhou Municipal Science and Technology Project under Grant 2024A04J6459.

\vspace{-0.1in}
\bibliographystyle{IEEEtran}
\bibliography{reference}

\section*{Biographies}

\footnotesize

\noindent\textbf{Zihao Tao} (ztao341@connect.hkust-gz.edu.cn) is currently pursuing his Ph.D. degree at the Hong Kong University of Science and Technology (Guangzhou), China. His research interests include wireless communications and sensing.
\medskip

\noindent\textbf{Yiming Zhao} (yzhao517@connect.hkust-gz.edu.cn) is currently pursuing his Ph.D. degree at the Hong Kong University of Science and Technology (Guangzhou), China. His research interests include multi-modal sensing.
\medskip

\noindent \textbf{Hongtao Zhao} (hzhao565@connect.hkust-gz.edu.cn) is currently pursuing his Ph.D. degree at the Hong Kong University of Science and Technology (Guangzhou), China. His research interests include integrated sensing and communication.
\medskip

\noindent \textbf{Zijun Gong} (gongzijun@hkust-gz.edu.cn) is currently an Assistant Professor at The Hong Kong University of Science and Technology (Guangzhou), China. 
His research interests include statistical signal processing and optimization, channel modeling, channel estimation, 
and localization.
\medskip

\noindent \textbf{Ying Cui} (yingcui@hkust-gz.edu.cn) is currently an Associate Professor at The Hong Kong University of Science and Technology (Guangzhou), China.
Her research interests include optimization, learning, IoT communications, integrated sensing and communications, and edge intelligence.
\end{document}